%% file: md-1.2.tex
\documentclass[11pt,twoside,dvips,letterpaper]{paper}

\usepackage{graphicx}
\usepackage{amsmath}
\usepackage{amssymb}
\usepackage{dcolumn}
\usepackage{multirow}
\usepackage{bm}
\usepackage[usenames,dvipsnames]{color}
\usepackage{mathrsfs}
\usepackage{alltt}
\usepackage[colorlinks,dvips]{hyperref}
\usepackage[hyphenbreaks]{breakurl}
\usepackage[margin=1in]{geometry}
\usepackage[compact]{titlesec}
\usepackage{mdwlist}

\newcommand{\bv}[1]{{\mbox{\boldmath$\mathbf{#1}$}}}
\newcommand{\irmean}[1]{{\langle{#1}\rangle}}
\newcommand{\irv}[1]{\langle{#1^2}\rangle}
\newcommand{\Eqr}[1]{(\ref{#1})}
\newcommand{\Eqre}[1]{Eq.~(\ref{#1})}
\newcommand{\Eqrs}[1]{(\ref{#1})}
\newcommand{\Eqres}[1]{Eqs.~(\ref{#1})}

\newcommand{\Fige}[1]{Figure~\ref{#1}}
\newcommand{\Figse}[1]{Figures~\ref{#1}}

\newcommand{\laur}{LA-UR-12-26980}

\newcommand{\ti}{A stochastic diffusion process for the Dirichlet distribution}
\newcommand{\aufirst}{J.\ Bakosi}
\newcommand{\ausecond}{J.R.\ Ristorcelli}
\newcommand{\au}{\aufirst and \ausecond}
\newcommand{\kw}{Fokker-Planck equation;
                 Stochastic diffusion;
                 Dirichlet distribution;
                 Monte-Carlo simulation}

\hypersetup{citecolor=blue,
			linkcolor=blue,
			urlcolor=blue,
			pdftitle=\ti,
			pdfauthor=\au,
			pdfkeywords=\kw}
\usepackage{hypernat}

\begin{document}

\title{\ti\\[0.2cm]\small\texttt{\laur}\\[0.2cm]\texttt{Accepted in
International Journal of Stochastic Analysis, March 1, 2013}\\[-0.3cm]}
\author{\normalsize\aufirst, \normalsize\ausecond\\[-0.1cm]
        \texttt{\normalsize\{jbakosi,jrrj\}@lanl.gov} \\[-0.1cm]
        \normalsize Los Alamos National Laboratory, Los Alamos, NM 87545,
        USA\\[-0.3cm]}

\maketitle

\begin{abstract}
The method of potential solutions of Fokker-Planck equations is used to develop
a transport equation for the joint probability of $N$ coupled stochastic
variables with the Dirichlet distribution as its asymptotic solution. To ensure
a bounded sample space, a coupled nonlinear diffusion process is required: the
Wiener-processes in the equivalent system of stochastic differential equations
are multiplicative with coefficients dependent on all the stochastic variables.
Individual samples of a discrete ensemble, obtained from the stochastic process,
satisfy a unit-sum constraint at all times. The process may be used to represent
realizations of a fluctuating ensemble of $N$ variables subject to a
conservation principle. Similar to the multivariate Wright-Fisher process, whose
invariant is also Dirichlet, the univariate case yields a process whose
invariant is the beta distribution. As a test of the results, Monte-Carlo
simulations are used to evolve numerical ensembles toward the invariant
Dirichlet distribution.
\keywords{\kw}
\end{abstract}

\section{Objective}
We develop a Fokker-Planck equation whose statistically stationary solution is
the Dirichlet distribution \cite{Johnson_60,Mosimann_62,Kotz_00}. The system of
stochastic differential equations (SDE), equivalent to the Fokker-Planck
equation, yields a Markov process that allows a Monte-Carlo method to
numerically evolve an ensemble of fluctuating variables that satisfy a unit-sum
requirement. A Monte Carlo solution is used to verify that the invariant
distribution is Dirichlet.

The Dirichlet distribution is a statistical representation of non-negative
variables subject to a unit-sum requirement. The properties of such variables
have been of interest in a variety of fields, including evolutionary theory
\cite{Pearson_1896}, Bayesian statistics \cite{Paulino_95}, geology
\cite{Chayes_62,Martin_65}, forensics \cite{Lange_95}, econometrics
\cite{Gourieroux_06}, turbulent combustion \cite{Girimaji_91}, and population
biology \cite{Steinrucken_2013}.

\section{Preview of results}
The Dirichlet distribution \cite{Johnson_60,Mosimann_62,Kotz_00} for a set of
scalars $0\!\le\!Y_\alpha$, $\alpha\!=\!1,\dots,N-1$,
$\sum_{\alpha=1}^{N-1}Y_\alpha\!\le\!1$, is given by
\begin{equation}
\mathscr{D}(\bv{Y},\bv{\omega}) =
\frac{\Gamma\left(\sum_{\alpha=1}^N\omega_\alpha\right)}
{\prod_{\alpha=1}^N\Gamma(\omega_\alpha)}\prod_{\alpha=1}^N
Y_\alpha^{\omega_\alpha-1},\label{eq:D}
\end{equation}
where $\omega_\alpha\!>\!0$ are parameters,
$Y_N\!=\!1-\sum_{\beta=1}^{N-1}Y_\beta$, and $\Gamma(\cdot)$ denotes the gamma
function. We derive the stochastic diffusion process, governing the scalars,
$Y_\alpha$,
\begin{equation}
\mathrm{d}Y_\alpha(t) = \frac{b_\alpha}{2}\big[S_\alpha Y_N -
(1-S_\alpha)Y_\alpha\big]\mathrm{d}t + \sqrt{\kappa_\alpha Y_\alpha
Y_N}\mathrm{d}W_\alpha(t), \qquad \alpha=1,\dots,N-1,\label{eq:iSDE}
\end{equation}
where $\mathrm{d}W_\alpha(t)$ is an isotropic vector-valued Wiener process
\cite{Gardiner_09}, and $b_\alpha\!>\!0$, $\kappa_\alpha\!>\!0$, and
$0\!<\!S_\alpha\!<\!1$ are coefficients. We show that the statistically
stationary solution of \Eqre{eq:iSDE} is the Dirichlet distribution,
\Eqre{eq:D}, provided the SDE coefficients satisfy
\begin{equation}
\frac{b_1}{\kappa_1}(1-S_1) = \dots = \frac{b_{N-1}}{\kappa_{N-1}}(1-S_{N-1}).
\end{equation}
The restrictions imposed on the SDE coefficients, $b_\alpha$, $\kappa_\alpha$,
and $S_\alpha$, ensure reflection towards the interior of the sample space,
which is a generalized triangle or tetrahedron (more precisely, a simplex) in
$N\!-\!1$ dimensions. The restrictions together with the specification of the
$N^\mathrm{th}$ scalar as $Y_N\!=\!1-\sum_{\beta=1}^{N-1}Y_\beta$ ensure
\begin{equation} \sum_{\alpha=1}^NY_\alpha=1.  \end{equation} Indeed, inspection
of \Eqre{eq:iSDE} shows that for example when $Y_1\!=\!0$, the diffusion is zero
and the drift is strictly positive, while if $Y_1\!=\!1$, the diffusion is zero
($Y_N\!=\!0$) and the drift is strictly negative.

\section{Development of the diffusion process}
\label{sec:DSDE}
The diffusion process \Eqr{eq:iSDE} is developed by the method of potential
solutions.

We start from the It\^o diffusion process \cite{Gardiner_09} for the stochastic
vector, $Y_\alpha$,
\begin{equation}
\mathrm{d}Y_\alpha(t) = a_\alpha(\bv{Y})\mathrm{d}t +
b_{\alpha\beta}(\bv{Y})\mathrm{d}W_\beta(t),
\qquad \alpha,\beta = 1,\dots,N-1,\label{eq:Ito}
\end{equation}
with drift, $a_\alpha(\bv{Y})$, diffusion, $b_{\alpha\beta}(\bv{Y})$, and the
isotropic vector-valued Wiener process, $\mathrm{d}W_\beta(t)$, where summation
is implied for repeated indices. Using standard methods given in
\cite{Gardiner_09} the equivalent Fokker-Planck equation governing the joint
probability, $\mathscr{F}(\bv{Y},t)$, derived from \Eqre{eq:Ito}, is
\begin{equation}
\frac{\partial\mathscr{F}}{\partial t} = -\frac{\partial}{\partial
Y_\alpha}\big[a_\alpha(\bv{Y})\mathscr{F}\big] +
\frac{1}{2}\frac{\partial^2}{\partial Y_\alpha \partial
Y_\beta}\big[B_{\alpha\beta}(\bv{Y})\mathscr{F}\big],\label{eq:FP}
\end{equation}
with diffusion $B_{\alpha\beta}\!=\!b_{\alpha\gamma}b_{\gamma\beta}$. Since the
drift and diffusion coefficients are time-homogeneous,
$a_\alpha(\bv{Y},t)\!=\!a_\alpha(\bv{Y})$ and
$B_{\alpha\beta}(\bv{Y},t)\!=\!B_{\alpha\beta}(\bv{Y})$, \Eqre{eq:Ito} is a
statistically stationary process and the solution of \Eqre{eq:FP} converges to a
stationary distribution, \cite{Gardiner_09} Sec.\ 6.2.2. Our task is to
specify the functional forms of $a_\alpha(\bv{Y})$ and $b_{\alpha\beta}(\bv{Y})$
so that the stationary solution of \Eqre{eq:FP} is $\mathscr{D}(\bv{Y})$,
defined by \Eqre{eq:D}.

A potential solution of \Eqre{eq:FP} exists if
\begin{equation}
\frac{\partial\ln\mathscr{F}}{\partial Y_\beta} =
B_{\alpha\beta}^{-1}\left(2a_\alpha - \frac{\partial B_{\alpha\gamma}}{\partial
Y_\gamma}\right) \equiv -\frac{\partial\phi}{\partial Y_\beta},
\qquad \alpha,\beta,\gamma = 1,\dots,N-1, \label{eq:solution}
\end{equation}
is satisfied, \cite{Gardiner_09} Sec.\ 6.2.2. Since the left hand side of
\Eqre{eq:solution} is a gradient, the expression on the right must also be a
gradient and can therefore be obtained from a scalar potential denoted by
$\phi(\bv{Y})$. This puts a constraint on the possible choices of $a_\alpha$ and
$B_{\alpha\beta}$ and on the potential, as
$\phi,_{\alpha\beta}=\phi,_{\beta\alpha}$ must also be satisfied. The potential
solution is
\begin{equation}
\mathscr{F}(\bv{Y}) = \exp[-\phi(\bv{Y})].\label{eq:phi}
\end{equation}
Now functional forms of $a_\alpha(\bv{Y})$ and $B_{\alpha\beta}(\bv{Y})$ that
satisfy \Eqre{eq:solution} with $\mathscr{F}(\bv{Y}) \equiv \mathscr{D}(\bv{Y})$
are sought. The mathematical constraints on the specification of $a_\alpha$ and
$B_{\alpha\beta}$ are as follows:
\begin{enumerate*}
\item $B_{\alpha\beta}$ must be symmetric positive semi-definite. This is to
ensure that
\begin{itemize*}
\item the square-root of $B_{\alpha\beta}$ (e.g.\ the Cholesky-decomposition,
$b_{\alpha\beta}$) exists, required by the correspondence of the SDE
\Eqr{eq:Ito} and the Fokker-Planck equation \Eqr{eq:FP},
\item \Eqre{eq:Ito} represents a diffusion, and
\item $\det(B_{\alpha\beta})\ne0$, required by the existence of the inverse in
\Eqre{eq:solution}.
\end{itemize*}
\item For a potential solution to exist \Eqre{eq:solution} must be satisfied.
\end{enumerate*}
With $\mathscr{F}(\bv{Y}) \equiv \mathscr{D}(\bv{Y})$ \Eqre{eq:phi} shows that
the scalar potential must be
\begin{equation}
\phi(\bv{Y}) = -\sum_{\alpha=1}^N(\omega_\alpha-1)\ln Y_\alpha.
\label{eq:potential}
\end{equation}
It is straightforward to verify that the specifications
{\allowdisplaybreaks
\begin{align}
a_\alpha(\bv{Y}) & = \frac{b_\alpha}{2}\big[S_\alpha Y_N -
(1-S_\alpha)Y_\alpha\big],\label{eq:a}\\\noalign{\smallskip}
B_{\alpha\beta}(\bv{Y}) & = \left\{
\begin{array}{lr}
\kappa_\alpha Y_\alpha Y_N & \quad \mathrm{for} \quad \alpha =
\beta,\\\noalign{\smallskip}
0 & \quad \mathrm{for} \quad \alpha \ne \beta,
\end{array}
\right.\label{eq:B}
\end{align}}%
satisfy the above mathematical constraints, 1.\ and 2. Here $b_\alpha\!>\!0$,
$\kappa_\alpha\!>\!0$, $0\!<\!S_\alpha\!<\!1$, and
$Y_N\!=\!1-\sum_{\beta=1}^{N-1}Y_\beta$. Summation is not implied for Eqs.\
(\ref{eq:potential}--\ref{eq:B}).

Substituting Eqs.\ (\ref{eq:potential}--\ref{eq:B}) into \Eqre{eq:solution}
yields a system with the same functions on both sides with different
coefficients, yielding the correspondence between the $N$ coefficients of the
Dirichlet distribution, \Eqre{eq:D}, and the Fokker-Planck equation
\Eqrs{eq:FP} with Eqs.\ (\ref{eq:a}--\ref{eq:B}) as
\begin{align}
\omega_\alpha & = \frac{b_\alpha}{\kappa_\alpha}S_\alpha, \qquad
\alpha=1,\dots,N-1,\label{eq:alphai}\\
\omega_N & = \frac{b_1}{\kappa_1}(1-S_1) = \dots =
\frac{b_{N-1}}{\kappa_{N-1}}(1-S_{N-1}).\label{eq:alphaN}
\end{align}%
For example, for $N\!=\!3$ one has $\bv{Y}=(Y_1,Y_2,Y_3=1-Y_1-Y_2)$ and from
\Eqre{eq:potential} the scalar potential is
\begin{equation}
-\phi(Y_1,Y_2) = (\omega_1-1)\ln Y_1 + (\omega_2-1)\ln Y_2
+ (\omega_3-1)\ln(1-Y_1-Y_2).
\end{equation}
\Eqre{eq:solution} then becomes the system
{\allowdisplaybreaks
\begin{align}
\frac{\omega_1-1}{Y_1} - \frac{\omega_3-1}{Y_3} & =
  \left(\frac{b_1}{\kappa_1}S_1-1\right)\frac{1}{Y_1} -
\left[\frac{b_1}{\kappa_1}(1-S_1)-1\right]\frac{1}{Y_3}, \\
\frac{\omega_2-1}{Y_2} - \frac{\omega_3-1}{Y_3} & =
  \left(\frac{b_2}{\kappa_2}S_2-1\right)\frac{1}{Y_2} -
\left[\frac{b_2}{\kappa_2}(1-S_2)-1\right]\frac{1}{Y_3},
\end{align}}%
which shows that by specifying the parameters, $\omega_\alpha$, of the
Dirichlet distribution as
{\allowdisplaybreaks
\begin{align}
\omega_1 & = \frac{b_1}{\kappa_1}S_1,\label{eq:alpha1}\\
\omega_2 & = \frac{b_2}{\kappa_2}S_2,\label{eq:alpha2}\\
\omega_3 & = \frac{b_1}{\kappa_1}(1-S_1) = \frac{b_2}{\kappa_2}(1-S_2),
\label{eq:alpha3}
\end{align}}%
the stationary solution of the Fokker-Planck equation \Eqrs{eq:FP} with drift
\Eqr{eq:a} and diffusion \Eqr{eq:B} is $\mathscr{D}(\bv{Y},\bv{\omega})$ for
$N\!=\!3$. The above development generalizes to $N$ variables, yielding Eqs.\
(\ref{eq:alphai}--\ref{eq:alphaN}), and reduces to the beta distribution, a
univariate specialization of $\mathscr{D}$ for $N\!=\!2$, where $Y_1\!=\!Y$ and
$Y_2\!=\!1-Y$, see \cite{Bakosi_beta}.

If Eqs.\ (\ref{eq:alphai}--\ref{eq:alphaN}) hold, the stationary solution of the
Fokker-Planck equation \Eqr{eq:FP} with drift \Eqr{eq:a} and diffusion
\Eqr{eq:B} is the Dirichlet distribution, \Eqre{eq:D}. Note that Eqs.\
(\ref{eq:a}--\ref{eq:B}) are one possible way of specifying a drift and a
diffusion to arrive at a Dirichlet distribution; other functional forms may be
possible. The specifications in Eqs.\ (\ref{eq:a}--\ref{eq:B}) are a
generalization of the results for a univariate diffusion process, discussed in
\cite{Bakosi_beta,Forman_08}, whose invariant distribution is beta.

The shape of the Dirichlet distribution, \Eqre{eq:D}, is determined by the $N$
coefficients, $\omega_\alpha$. Eqs.\ (\ref{eq:alphai}--\ref{eq:alphaN}) show
that in the stochastic system, different combinations of $b_\alpha$,
$S_\alpha$, and $\kappa_\alpha$ may yield the same $\omega_\alpha$ and that not
all of $b_\alpha$, $S_\alpha$, and $\kappa_\alpha$ may be chosen independently
to keep the invariant Dirichlet.

\section{Corroborating that the invariant distribution is Dirichlet}
\label{sec:numerics}
For any multivariate Fokker-Planck equation there is an equivalent system of
It\^o diffusion processes, such as the pair of Eqs.\ (\ref{eq:Ito}--\ref{eq:FP})
\cite{Gardiner_09}. Therefore, a way of computing the (discrete) numerical
solution of \Eqre{eq:FP} is to integrate \Eqre{eq:Ito} in a Monte-Carlo fashion
for an ensemble \cite{Pope_85}. Using a Monte-Carlo simulation we show that the
statistically stationary solution of the Fokker-Planck equation \Eqr{eq:FP} with
drift and diffusion (\ref{eq:a}--\ref{eq:B}) is a Dirichlet distribution,
\Eqre{eq:D}.

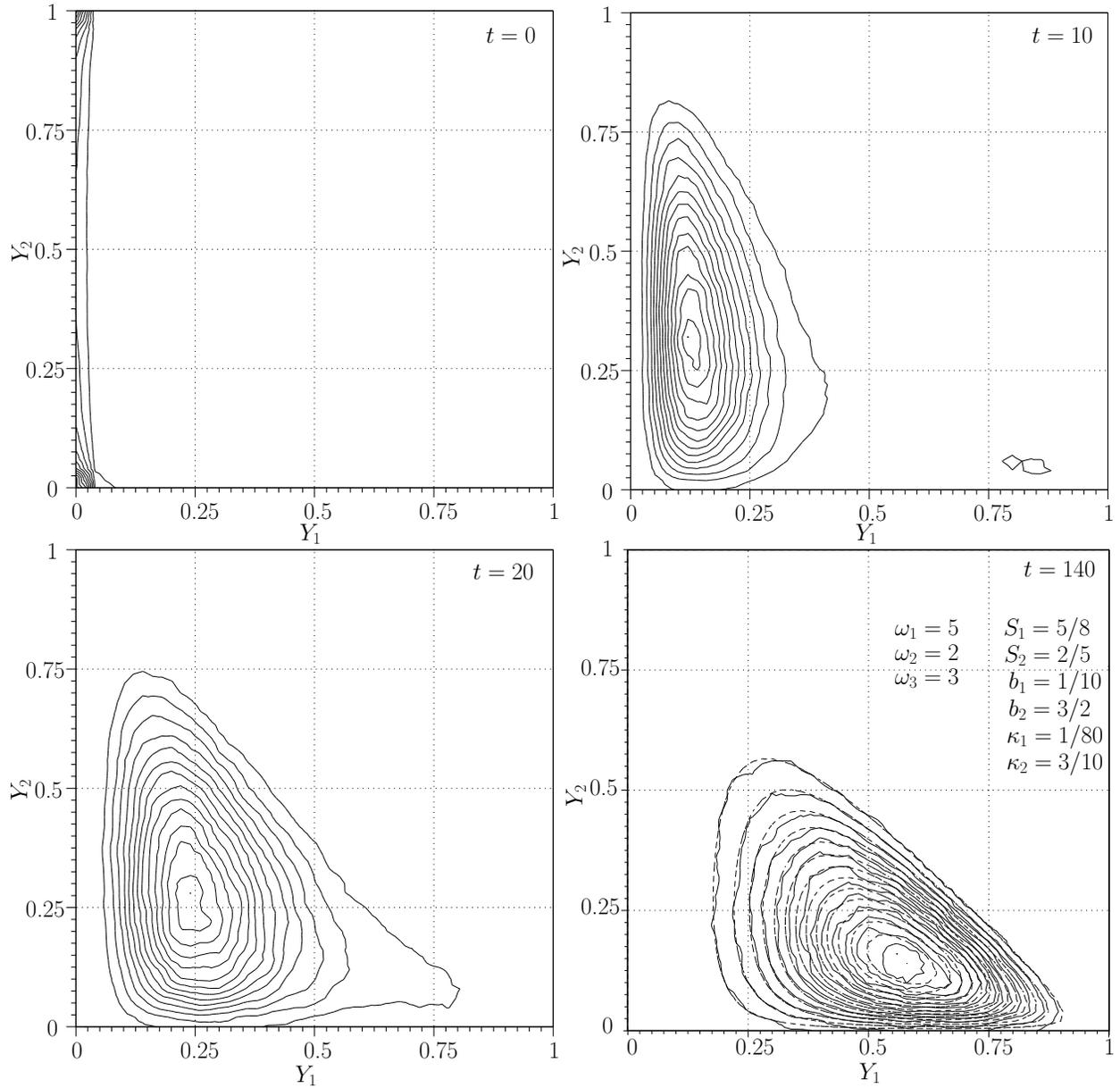
\begin{figure}
\centering
\resizebox{1.0\columnwidth}{!}{\input{jpdfs.pstex_t}}
\caption{Time evolution of the joint probability, $\mathscr{F}(Y_1,Y_2)$,
extracted from the numerical solution of Eqs. (\ref{eq:Ito1}--\ref{eq:Ito3}).
The initial condition is a triple-delta distribution, with unequal peaks at the
three corners of the sample space. At the end of the simulation, $t=140$, the
solid lines are that of the distribution extracted from the numerical
ensemble, dashed lines are that of a Dirichlet distribution to which the
solution converges in the statistically stationary state, implied by the
constant SDE coefficients, sampled at the same heights.}
\label{fig:jpdfs}
\end{figure}

\begin{figure}
\centering
\resizebox{1.0\columnwidth}{!}{\input{jpdfs_box.pstex_t}}
\caption{Time evolution of the joint probability, $\mathscr{F}(Y_1,Y_2)$,
extracted from the numerical solution of Eqs. (\ref{eq:Ito1}--\ref{eq:Ito3}).
The top-left panel shows the initial condition: a box with diffused sides. By
$t=160$, bottom-right panel, the distribution converges to the same Dirichlet
distribution as in \Fige{fig:jpdfs}.}
\label{fig:jpdfs_box}
\end{figure}
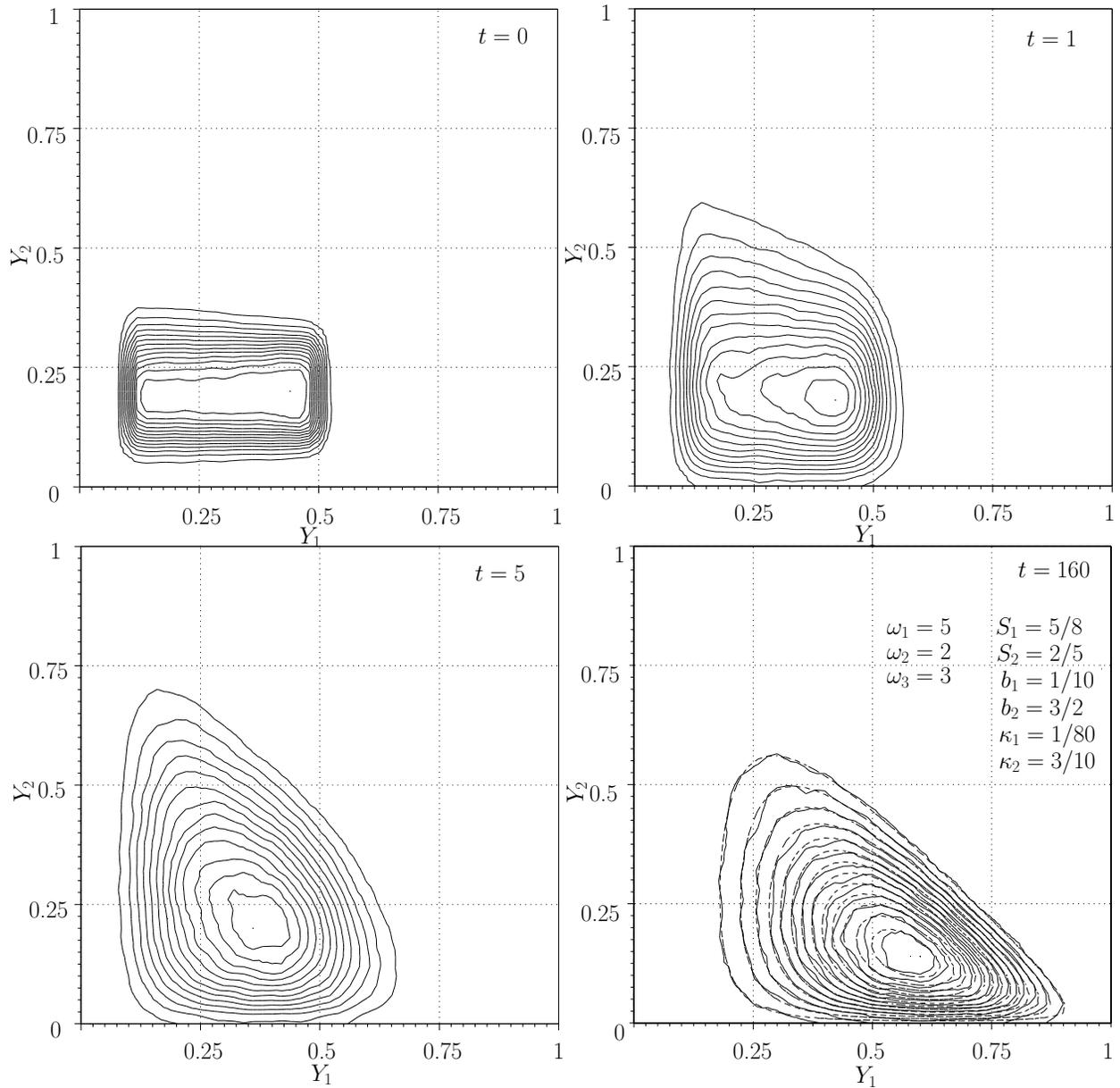

The time-evolution of an ensemble of particles, each with $N=3$ variables
($Y_1,Y_2,Y_3$), is numerically computed by integrating the system of equations
\Eqr{eq:Ito}, with drift and diffusion (\ref{eq:a}--\ref{eq:B}), for $N=3$
as
\begin{align}
&\hspace{-3cm}\qquad \mathrm{d}Y^{(i)}_1 = \frac{b_1}{2}\left[S_1
  Y^{(i)}_3 - (1-S_1)Y^{(i)}_1\right]\mathrm{d}t + \sqrt{\kappa_1
  Y^{(i)}_1 Y^{(i)}_3}\mathrm{d}W^{(i)}_1 \label{eq:Ito1}\\
&\hspace{-3cm}\qquad \mathrm{d}Y^{(i)}_2 = \frac{b_2}{2}\left[S_2
  Y^{(i)}_3 - (1-S_2)Y^{(i)}_2\right]\mathrm{d}t + \sqrt{\kappa_2
  Y^{(i)}_2 Y^{(i)}_3}\mathrm{d}W^{(i)}_2 \label{eq:Ito2}\\
&\hspace{-3cm}\enskip\qquad Y^{(i)}_3 =
  1-Y^{(i)}_1-Y^{(i)}_2 \label{eq:Ito3}
\end{align}%
for each particle $i$. In Eqs.\ (\ref{eq:Ito1}--\ref{eq:Ito2}) $\mathrm{d}W_1$
and $\mathrm{d}W_2$ are independent Wiener processes, sampled from Gaussian
streams of random numbers with mean $\irmean{\mathrm{d}W_\alpha}\!=\!0$ and
covariance $\irmean{\mathrm{d}W_\alpha\mathrm{d}W_\beta}\!=
\!\delta_{\alpha\beta}\mathrm{d}t$. $400,\!000$ particle-triplets,
$(Y_1,Y_2,Y_3)$, are generated with two different initial distributions,
displayed in the upper-left of \Figse{fig:jpdfs} and \ref{fig:jpdfs_box}, a
triple-delta and a box, respectively. Each member of both initial ensembles
satisfy $\sum_{\alpha=1}^3Y_\alpha\!=\!1$. Eqs.\ (\ref{eq:Ito1}--\ref{eq:Ito3})
are advanced in time with the Euler-Maruyama scheme \cite{Kloeden_99} with time
step $\Delta t\!=\!0.05$. Table \ref{tab:ic} shows the coefficients of the
stochastic system (\ref{eq:Ito1}--\ref{eq:Ito3}), the corresponding parameters
of the final Dirichlet distribution, and the first two moments at the initial
times for the triple-delta initial condition case. The final state of the
ensembles are determined by the SDE coefficients, constant for these exercises,
also given in Table \ref{tab:ic}, the same for both simulations, satisfying
\Eqre{eq:alpha3}.

The time-evolutions of the joint probabilities are extracted from both
calculations and displayed at different times in \Figse{fig:jpdfs} and
\ref{fig:jpdfs_box}. At the end of the simulations two distributions are
plotted at the bottom-right of both figures: the one extracted from the
numerical ensemble and the Dirichlet distribution determined analytically using
the SDE coefficients -- in excellent agreement in both figures. The
statistically stationary solution of the developed stochastic system is the
Dirichlet distribution.

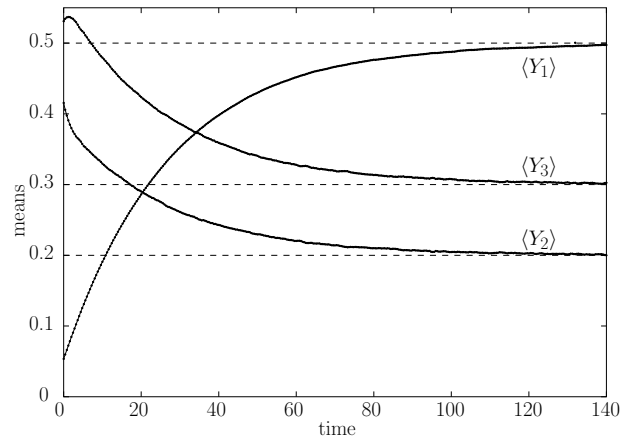
\begin{figure}
\centering
\resizebox{0.5\columnwidth}{!}{\input{means.pstex_t}}
\caption{Time evolution of the means extracted from the numerically integrated
system of Eqs.\ (\ref{eq:Ito1}--\ref{eq:Ito3}) starting from the triple-delta
initial condition. Dotted-solid lines -- numerical solution, dashed lines --
statistics of the Dirichlet distribution determined analytically using the
constant coefficients of the SDE, see Table \ref{tab:ic}.}
\label{fig:mean}
\end{figure}

\begin{figure}
\centering
\resizebox{1.0\columnwidth}{!}{\input{secmom.pstex_t}}
\caption{Time evolution of the second central moments extracted from the
numerically integrated system of Eqs.\ (\ref{eq:Ito1}--\ref{eq:Ito3}) starting
from the triple-delta initial condition. The legend is the same as in
\Fige{fig:mean}.}
\label{fig:secmom}
\end{figure}
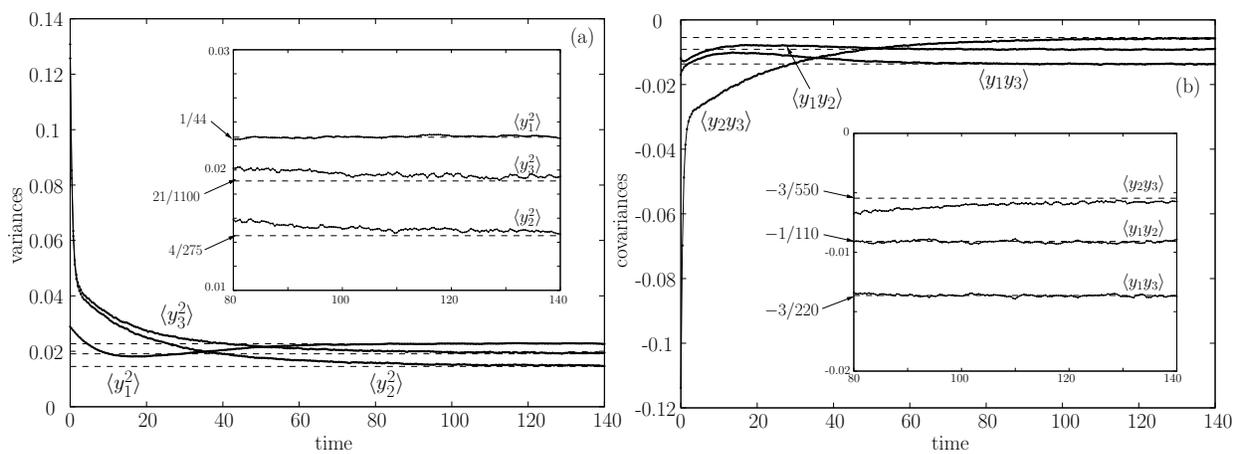

For a more quantitative evaluation, the time evolution of the first two
moments, $\mu_\alpha\!=\!\irmean{Y_\alpha}\!=\!\int_0^1\int_0^1 Y_\alpha
\mathscr{F}(Y_1,Y_2) \mathrm{d}Y_1 \mathrm{d}Y_2$, and $\irmean{y_\alpha
y_\beta}\!=\!\irmean{(Y_\alpha\!-\!\irmean{Y_\alpha})
(Y_\beta\!-\!\irmean{Y_\beta})}$,
are also extracted from the numerical simulation with the triple-delta-peak
initial condition as ensemble averages and displayed in \Figse{fig:mean} and
\ref{fig:secmom}. The figures show that the statistics converge to the precise
state given by the Dirichlet distribution that is prescribed by the SDE
coefficients, see Table \ref{tab:ic}.

The solution approaches a Dirichlet distribution, with non-positive covariances
\cite{Mosimann_62}, in the statistically stationary limit, \Fige{fig:secmom}(b).
Note that during the evolution of the process, $0\!<\!t\!\lesssim\!80$, the
solution is not necessarily Dirichlet, but the stochastic variables sum to one
at all times. The point ($Y_1$, $Y_2$), governed by Eqs.\
(\ref{eq:Ito1}--\ref{eq:Ito2}), can never leave the $(N\!-\!1)$-dimensional
(here $N\!=\!3$) convex polytope and by definition $Y_3\!=\!1-Y_1-Y_2$. The rate
at which the numerical solution converges to a Dirichlet distribution is
determined by the vectors $b_\alpha$ and $\kappa_\alpha$.

The above numerical results confirm that starting from arbitrary realizable
ensembles the solution of the stochastic system converges to a Dirichlet
distribution in the statistically stationary state, specified by the SDE
coefficients.

\section{Relation to other diffusion processes}
\label{sec:relation}
It is useful to relate the Dirichlet diffusion process, \Eqre{eq:iSDE}, to other
multivariate stochastic diffusion processes with linear drift and quadratic
diffusion.

A close relative of \Eqre{eq:iSDE} is the multivariate Wright-Fisher (WF)
process \cite{Steinrucken_2013}, used extensively in population and genetic
biology,
\begin{equation}
\mathrm{d}Y_\alpha(t) = \frac{1}{2} (\omega_\alpha-\omega Y_\alpha) \mathrm{d}t
+ \sum_{\beta=1}^{N-1} \sqrt{Y_\alpha(\delta_{\alpha\beta}-Y_\beta)}
\mathrm{d}W_{\alpha\beta}(t), \qquad \alpha = 1,\dots,N-1,
\label{eq:WF}
\end{equation}
where $\delta_{\alpha\beta}$ is Kronecker's delta,
$\omega\!=\!\sum_{\beta=1}^N\omega_\beta$ with $\omega_\alpha$ defined in
\Eqre{eq:D} and, $Y_N\!=\!1\!-\!\sum_{\beta=1}^{N-1}Y_\beta$. Similarly to
\Eqre{eq:iSDE}, the statistically stationary solution of \Eqre{eq:WF} is the
Dirichlet distribution \cite{Karlin_1981}. It is straightforward to verify that
its drift and diffusion also satisfy \Eqre{eq:solution} with
$\mathscr{F}\equiv\mathscr{D}$, i.e.\ WF is a process whose invariant is
Dirichlet and this solution is potential. A notable difference between
\Eqres{eq:iSDE} and \Eqr{eq:WF}, other than the coefficients, is that the
diffusion matrix of the Dirichlet diffusion process is diagonal, while that of
the WF process it is full.

Another process similar to \Eqres{eq:iSDE} and \Eqr{eq:WF} is the multivariate
Jacobi process, used in econometrics,
\begin{equation}
\mathrm{d}Y_\alpha(t) = a(Y_\alpha-\pi_\alpha)\mathrm{d}t +
\sqrt{cY_\alpha}\mathrm{d}W_\alpha(t) - \sum_{\beta=1}^{N-1}
Y_\alpha\sqrt{cY_\beta}\mathrm{d}W_\beta(t), \qquad \alpha = 1,\dots,N
\label{eq:Jacobi}
\end{equation}
of Gourieroux \& Jasiak \cite{Gourieroux_06} with $a<0$, $c>0$, $\pi_\alpha>0$,
and $\sum_{\beta=1}^N\pi_\beta=1$.

In the univariate case the Dirichlet, WF, and Jacobi diffusions reduce to
\begin{equation}
\mathrm{d}Y(t) = \frac{b}{2} (S-Y)\mathrm{d}t +
                 \sqrt{\kappa Y(1-Y)}\mathrm{d}W(t),
\label{eq:beta}
\end{equation}
see also \cite{Bakosi_beta}, whose invariant is the beta distribution, which
belongs to the family of Pearson diffusions, discussed in detail by Forman \&
Sorensen \cite{Forman_08}.

\section{Summary}
\label{sec:summary}
The method of potential solutions of Fokker-Planck equations has been used to
derive a transport equation for the joint distribution of $N$ fluctuating
variables. The equivalent stochastic process, governing the set of random
variables, $0\!\le\!Y_\alpha$, $\alpha\!=\!1,\dots,N-1$,
$\sum_{\alpha=1}^{N-1}Y_\alpha\!\le\!1$, reads
\begin{equation}
\mathrm{d}Y_\alpha(t) = \frac{b_\alpha}{2} \big[S_\alpha Y_N -
(1-S_\alpha)Y_\alpha\big] \mathrm{d}t + \sqrt{\kappa_\alpha Y_\alpha Y_N}
\mathrm{d}W_\alpha(t), \qquad \alpha=1,\dots,N-1,
\label{eq:sumSDE}
\end{equation}
where $Y_N\!=\!1\!-\!\sum_{\beta=1}^{N-1}Y_\beta$, and $b_\alpha$,
$\kappa_\alpha$ and $S_\alpha$ are parameters, while $\mathrm{d}W_\alpha(t)$ is
an isotropic Wiener process with independent increments. Restricting the
coefficients to $b_\alpha\!>\!0$, $\kappa_\alpha\!>\!0$ and
$0\!<\!S_\alpha\!<\!1$, and defining $Y_N$ as above ensure
$\sum_{\alpha=1}^NY_\alpha\!=\!1$ and that individual realizations of
($Y_1,Y_2,\dots,Y_N)$ are confined to the ($N\!-\!1$)-dimensional convex
polytope of the sample space. \Eqre{eq:sumSDE} can therefore be used to
numerically evolve the joint distribution of $N$ fluctuating variables required
to satisfy a conservation principle. \Eqre{eq:sumSDE} is a coupled system of
nonlinear stochastic differential equations whose statistically stationary
solution is the Dirichlet distribution, \Eqre{eq:D}, provided the coefficients
satisfy
\begin{equation}
\frac{b_1}{\kappa_1}(1-S_1) = \dots = \frac{b_{N-1}}{\kappa_{N-1}}(1-S_{N-1}).
\end{equation}

In stochastic modeling, one typically begins with a physical problem, perhaps
discrete, \emph{then} derives the stochastic differential equations whose
solution yields a distribution. In this paper we reversed the process: we
assumed a desired stationary distribution and derived the stochastic
differential equations that converge to the assumed distribution. A potential
solution form of the Fokker-Planck equation was posited, from which we obtained
the stochastic differential equations for the diffusion process whose
statistically stationary solution is the Dirichlet distribution. We have also
made connections to other stochastic processes, such as the Wright-Fisher
diffusions of population biology and the Jacobi diffusions in econometrics,
whose invariant distributions possess similar properties but whose stochastic
differential equations are different.

\section*{Acknowledgements}
It is a pleasure to acknowledge a series of informative discussions with J.\
Waltz. This work was performed under the auspices of the U.S.\ Department of
Energy under the Advanced Simulation and Computing Program.

\newpage

\bibliographystyle{unsrt}
\bibliography{jbakosi}

\begin{table}
\setlength{\abovecaptionskip}{0pt}
\setlength{\belowcaptionskip}{-0.2cm}
\caption{\label{tab:ic}Initial and final states of the Monte-Carlo simulation
starting from a triple-delta. The coefficients, $b_1$, $b_2$, $S_1$, $S_2$,
$\kappa_1$, $\kappa_2$, of the system of SDEs (\ref{eq:Ito1}--\ref{eq:Ito3})
determine the distribution to which the system converges. The Dirichlet
parameters, implied by the SDE coefficients via Eqs.\
(\ref{eq:alpha1}-\ref{eq:alpha3}), are in brackets. The corresponding statistics
are determined by the well-known formulae of Dirichlet distributions
\cite{Mosimann_62}.}
\begin{center}
\begin{tabular}{r@{\hspace{0.1cm}}c@{\hspace{0.1cm}}lr@{\hspace{0.1cm}}
c@{\hspace{0.1cm}}l|r@{\hspace{0.1cm}}c@{\hspace{0.1cm}}l@{\hspace{-0.2cm}}
r@{\hspace{0.1cm}}c@{\hspace{0.1cm}}l@{\hspace{-0.3cm}}r@{\hspace{0.1cm}}
c@{\hspace{0.1cm}}l}
\multicolumn{6}{p{0.27\columnwidth}}{Initial state: triple-delta, see
\Fige{fig:jpdfs}} &
\multicolumn{9}{p{0.63\columnwidth}}{SDE coefficients and the
statistics of their implied Dirichlet distribution in the stationary state} \\
\hline
&& && && $b_1$&=&$1/10$ & $b_2$&=&$3/2$ & $(\omega_1$&=&$5)$\\[0.1cm]
&& && && $S_1$&=&$5/8$ & $S_2$&=&$2/5$ & $(\omega_2$&=&$2)$\\[0.1cm]
&& && && $\kappa_1$&=&$1/80$ & $\kappa_2$&=&$3/10$&
$\qquad(\omega_3$&=&$3)$\\[0.1cm]
$\irmean{Y_1}_0$&$\approx$&$0.05$ & && & $\irmean{Y_1}_\mathrm{s}$&=&$1/2$ & &&
& && \\[0.1cm]
$\irmean{Y_2}_0$&$\approx$&$0.42$ & && & $\irmean{Y_2}_\mathrm{s}$&=&$1/5$ & &&
& && \\[0.1cm]
$\irmean{Y_3}_0$&$\approx$&$0.53$ & && & $\irmean{Y_3}_\mathrm{s}$&=&$3/10$ &
&& & && \\[0.1cm]
$\irv{y_1}_0$&$\approx$&$0.03$ & && & $\irv{y_1}_\mathrm{s}$&=&$1/44$ & && & &&
\\[0.1cm]
$\irv{y_2}_0$&$\approx$&$0.125$ & && & $\irv{y_2}_\mathrm{s}$&=&$4/275$ & && &
&& \\[0.1cm]
$\irv{y_3}_0$&$\approx$&$0.13$ & && & $\irv{y_3}_\mathrm{s}$&=&$21/1100$ & && &
&& \\[0.1cm]
$\irmean{y_1y_2}_0$&$\approx$&$-0.012$ & && &
$\irmean{y_1y_2}_\mathrm{s}$&=&$-1/110$ & && & && \\[0.1cm]
$\irmean{y_1y_3}_0$&$\approx$&$-0.017$ & && &
$\irmean{y_1y_3}_\mathrm{s}$&=&$-3/220$ & && & && \\[0.1cm]
$\irmean{y_2y_3}_0$&$\approx$&$-0.114$ & && &
$\irmean{y_2y_3}_\mathrm{s}$&=&$-3/550$
\end{tabular}
\end{center}
\end{table}

\end{document}

%% file: jpdfs.pstex_t
\begin{picture}(0,0)%
\includegraphics{jpdfs.pstex}%
\end{picture}%
\setlength{\unitlength}{3947sp}%
\begingroup\makeatletter\ifx\SetFigFont\undefined%
\gdef\SetFigFont#1#2#3#4#5{%
  \reset@font\fontsize{#1}{#2pt}%
  \fontfamily{#3}\fontseries{#4}\fontshape{#5}%
  \selectfont}%
\fi\endgroup%
\begin{picture}(19195,18815)(-379,-18038)
%  text 
\put(4651,-8536){\makebox(0,0)[lb]{\smash{{\SetFigFont{25}{30.0}{\familydefault}{\mddefault}{\updefault}{\color[rgb]{0,0,0}$Y_1$}%
}}}}
%  text 
\put(4576,-17911){\makebox(0,0)[lb]{\smash{{\SetFigFont{25}{30.0}{\familydefault}{\mddefault}{\updefault}{\color[rgb]{0,0,0}$Y_1$}%
}}}}
%  text 
\put(17326, 89){\makebox(0,0)[lb]{\smash{{\SetFigFont{25}{30.0}{\familydefault}{\mddefault}{\updefault}{\color[rgb]{0,0,0}$t=10$}%
}}}}
%  text 
\put(7876, 89){\makebox(0,0)[lb]{\smash{{\SetFigFont{25}{30.0}{\familydefault}{\mddefault}{\updefault}{\color[rgb]{0,0,0}$t=0$}%
}}}}
%  text 
\put(7651,-9211){\makebox(0,0)[lb]{\smash{{\SetFigFont{25}{30.0}{\familydefault}{\mddefault}{\updefault}{\color[rgb]{0,0,0}$t=20$}%
}}}}
%  text 
\put(704,-8157){\makebox(0,0)[lb]{\smash{{\SetFigFont{25}{30.0}{\familydefault}{\mddefault}{\updefault}{\color[rgb]{0,0,0}0}%
}}}}
%  text 
\put(2586,-8157){\makebox(0,0)[lb]{\smash{{\SetFigFont{25}{30.0}{\familydefault}{\mddefault}{\updefault}{\color[rgb]{0,0,0}0.25}%
}}}}
%  text 
\put(4726,-8157){\makebox(0,0)[lb]{\smash{{\SetFigFont{25}{30.0}{\familydefault}{\mddefault}{\updefault}{\color[rgb]{0,0,0}0.5}%
}}}}
%  text 
\put(6716,-8157){\makebox(0,0)[lb]{\smash{{\SetFigFont{25}{30.0}{\familydefault}{\mddefault}{\updefault}{\color[rgb]{0,0,0}0.75}%
}}}}
%  text 
\put(8964,-8157){\makebox(0,0)[lb]{\smash{{\SetFigFont{25}{30.0}{\familydefault}{\mddefault}{\updefault}{\color[rgb]{0,0,0}1}%
}}}}
%  text 
\put(704,-17491){\makebox(0,0)[lb]{\smash{{\SetFigFont{25}{30.0}{\familydefault}{\mddefault}{\updefault}{\color[rgb]{0,0,0}0}%
}}}}
%  text 
\put(2586,-17491){\makebox(0,0)[lb]{\smash{{\SetFigFont{25}{30.0}{\familydefault}{\mddefault}{\updefault}{\color[rgb]{0,0,0}0.25}%
}}}}
%  text 
\put(4726,-17491){\makebox(0,0)[lb]{\smash{{\SetFigFont{25}{30.0}{\familydefault}{\mddefault}{\updefault}{\color[rgb]{0,0,0}0.5}%
}}}}
%  text 
\put(6716,-17491){\makebox(0,0)[lb]{\smash{{\SetFigFont{25}{30.0}{\familydefault}{\mddefault}{\updefault}{\color[rgb]{0,0,0}0.75}%
}}}}
%  text 
\put(8964,-17491){\makebox(0,0)[lb]{\smash{{\SetFigFont{25}{30.0}{\familydefault}{\mddefault}{\updefault}{\color[rgb]{0,0,0}1}%
}}}}
%  text 
\put(10304,-8191){\makebox(0,0)[lb]{\smash{{\SetFigFont{25}{30.0}{\familydefault}{\mddefault}{\updefault}{\color[rgb]{0,0,0}0}%
}}}}
%  text 
\put(12186,-8191){\makebox(0,0)[lb]{\smash{{\SetFigFont{25}{30.0}{\familydefault}{\mddefault}{\updefault}{\color[rgb]{0,0,0}0.25}%
}}}}
%  text 
\put(14326,-8191){\makebox(0,0)[lb]{\smash{{\SetFigFont{25}{30.0}{\familydefault}{\mddefault}{\updefault}{\color[rgb]{0,0,0}0.5}%
}}}}
%  text 
\put(16316,-8191){\makebox(0,0)[lb]{\smash{{\SetFigFont{25}{30.0}{\familydefault}{\mddefault}{\updefault}{\color[rgb]{0,0,0}0.75}%
}}}}
%  text 
\put(18564,-8191){\makebox(0,0)[lb]{\smash{{\SetFigFont{25}{30.0}{\familydefault}{\mddefault}{\updefault}{\color[rgb]{0,0,0}1}%
}}}}
%  text 
\put(14273,-8536){\makebox(0,0)[lb]{\smash{{\SetFigFont{25}{30.0}{\familydefault}{\mddefault}{\updefault}{\color[rgb]{0,0,0}$Y_1$}%
}}}}
%  text 
\put(9492,-3811){\rotatebox{90.0}{\makebox(0,0)[lb]{\smash{{\SetFigFont{25}{30.0}{\familydefault}{\mddefault}{\updefault}{\color[rgb]{0,0,0}$Y_2$}%
}}}}}
%  text 
\put(-49,-3720){\rotatebox{90.0}{\makebox(0,0)[lb]{\smash{{\SetFigFont{25}{30.0}{\familydefault}{\mddefault}{\updefault}{\color[rgb]{0,0,0}$Y_2$}%
}}}}}
%  text 
\put(-65,-13036){\rotatebox{90.0}{\makebox(0,0)[lb]{\smash{{\SetFigFont{25}{30.0}{\familydefault}{\mddefault}{\updefault}{\color[rgb]{0,0,0}$Y_2$}%
}}}}}
%  text 
\put(280,-7783){\makebox(0,0)[lb]{\smash{{\SetFigFont{25}{30.0}{\familydefault}{\mddefault}{\updefault}{\color[rgb]{0,0,0}0}%
}}}}
%  text 
\put(-87,-5718){\makebox(0,0)[lb]{\smash{{\SetFigFont{25}{30.0}{\familydefault}{\mddefault}{\updefault}{\color[rgb]{0,0,0}0.25}%
}}}}
%  text 
\put( 63,-3653){\makebox(0,0)[lb]{\smash{{\SetFigFont{25}{30.0}{\familydefault}{\mddefault}{\updefault}{\color[rgb]{0,0,0}0.5}%
}}}}
%  text 
\put(-87,-1588){\makebox(0,0)[lb]{\smash{{\SetFigFont{25}{30.0}{\familydefault}{\mddefault}{\updefault}{\color[rgb]{0,0,0}0.75}%
}}}}
%  text 
\put(280,477){\makebox(0,0)[lb]{\smash{{\SetFigFont{25}{30.0}{\familydefault}{\mddefault}{\updefault}{\color[rgb]{0,0,0}1}%
}}}}
%  text 
\put(280,-17117){\makebox(0,0)[lb]{\smash{{\SetFigFont{25}{30.0}{\familydefault}{\mddefault}{\updefault}{\color[rgb]{0,0,0}0}%
}}}}
%  text 
\put(-87,-15052){\makebox(0,0)[lb]{\smash{{\SetFigFont{25}{30.0}{\familydefault}{\mddefault}{\updefault}{\color[rgb]{0,0,0}0.25}%
}}}}
%  text 
\put( 63,-12987){\makebox(0,0)[lb]{\smash{{\SetFigFont{25}{30.0}{\familydefault}{\mddefault}{\updefault}{\color[rgb]{0,0,0}0.5}%
}}}}
%  text 
\put(-87,-10922){\makebox(0,0)[lb]{\smash{{\SetFigFont{25}{30.0}{\familydefault}{\mddefault}{\updefault}{\color[rgb]{0,0,0}0.75}%
}}}}
%  text 
\put(280,-8857){\makebox(0,0)[lb]{\smash{{\SetFigFont{25}{30.0}{\familydefault}{\mddefault}{\updefault}{\color[rgb]{0,0,0}1}%
}}}}
%  text 
\put(9880,-7817){\makebox(0,0)[lb]{\smash{{\SetFigFont{25}{30.0}{\familydefault}{\mddefault}{\updefault}{\color[rgb]{0,0,0}0}%
}}}}
%  text 
\put(9513,-5752){\makebox(0,0)[lb]{\smash{{\SetFigFont{25}{30.0}{\familydefault}{\mddefault}{\updefault}{\color[rgb]{0,0,0}0.25}%
}}}}
%  text 
\put(9663,-3687){\makebox(0,0)[lb]{\smash{{\SetFigFont{25}{30.0}{\familydefault}{\mddefault}{\updefault}{\color[rgb]{0,0,0}0.5}%
}}}}
%  text 
\put(9513,-1622){\makebox(0,0)[lb]{\smash{{\SetFigFont{25}{30.0}{\familydefault}{\mddefault}{\updefault}{\color[rgb]{0,0,0}0.75}%
}}}}
%  text 
\put(9880,443){\makebox(0,0)[lb]{\smash{{\SetFigFont{25}{30.0}{\familydefault}{\mddefault}{\updefault}{\color[rgb]{0,0,0}1}%
}}}}
%  text 
\put(14333,-17864){\makebox(0,0)[lb]{\smash{{\SetFigFont{25}{30.0}{\familydefault}{\mddefault}{\updefault}{\color[rgb]{0,0,0}$Y_1$}%
}}}}
%  text 
\put(17183,-9164){\makebox(0,0)[lb]{\smash{{\SetFigFont{25}{30.0}{\familydefault}{\mddefault}{\updefault}{\color[rgb]{0,0,0}$t=140$}%
}}}}
%  text 
\put(10311,-17444){\makebox(0,0)[lb]{\smash{{\SetFigFont{25}{30.0}{\familydefault}{\mddefault}{\updefault}{\color[rgb]{0,0,0}0}%
}}}}
%  text 
\put(12193,-17444){\makebox(0,0)[lb]{\smash{{\SetFigFont{25}{30.0}{\familydefault}{\mddefault}{\updefault}{\color[rgb]{0,0,0}0.25}%
}}}}
%  text 
\put(14333,-17444){\makebox(0,0)[lb]{\smash{{\SetFigFont{25}{30.0}{\familydefault}{\mddefault}{\updefault}{\color[rgb]{0,0,0}0.5}%
}}}}
%  text 
\put(16323,-17444){\makebox(0,0)[lb]{\smash{{\SetFigFont{25}{30.0}{\familydefault}{\mddefault}{\updefault}{\color[rgb]{0,0,0}0.75}%
}}}}
%  text 
\put(18571,-17444){\makebox(0,0)[lb]{\smash{{\SetFigFont{25}{30.0}{\familydefault}{\mddefault}{\updefault}{\color[rgb]{0,0,0}1}%
}}}}
%  text 
\put(9542,-13139){\rotatebox{90.0}{\makebox(0,0)[lb]{\smash{{\SetFigFont{25}{30.0}{\familydefault}{\mddefault}{\updefault}{\color[rgb]{0,0,0}$Y_2$}%
}}}}}
%  text 
\put(9887,-17070){\makebox(0,0)[lb]{\smash{{\SetFigFont{25}{30.0}{\familydefault}{\mddefault}{\updefault}{\color[rgb]{0,0,0}0}%
}}}}
%  text 
\put(9520,-15005){\makebox(0,0)[lb]{\smash{{\SetFigFont{25}{30.0}{\familydefault}{\mddefault}{\updefault}{\color[rgb]{0,0,0}0.25}%
}}}}
%  text 
\put(9670,-12940){\makebox(0,0)[lb]{\smash{{\SetFigFont{25}{30.0}{\familydefault}{\mddefault}{\updefault}{\color[rgb]{0,0,0}0.5}%
}}}}
%  text 
\put(9520,-10875){\makebox(0,0)[lb]{\smash{{\SetFigFont{25}{30.0}{\familydefault}{\mddefault}{\updefault}{\color[rgb]{0,0,0}0.75}%
}}}}
%  text 
\put(9887,-8810){\makebox(0,0)[lb]{\smash{{\SetFigFont{25}{30.0}{\familydefault}{\mddefault}{\updefault}{\color[rgb]{0,0,0}1}%
}}}}
%  text 
\put(14954,-10181){\makebox(0,0)[lb]{\smash{{\SetFigFont{25}{30.0}{\familydefault}{\mddefault}{\updefault}{\color[rgb]{0,0,0}$\omega_1=5$}%
}}}}
%  text 
\put(14954,-10606){\makebox(0,0)[lb]{\smash{{\SetFigFont{25}{30.0}{\familydefault}{\mddefault}{\updefault}{\color[rgb]{0,0,0}$\omega_2=2$}%
}}}}
%  text 
\put(14954,-11030){\makebox(0,0)[lb]{\smash{{\SetFigFont{25}{30.0}{\familydefault}{\mddefault}{\updefault}{\color[rgb]{0,0,0}$\omega_3=3$}%
}}}}
%  text 
\put(16832,-10180){\makebox(0,0)[lb]{\smash{{\SetFigFont{25}{30.0}{\familydefault}{\mddefault}{\updefault}{\color[rgb]{0,0,0}$S_1=5/8$}%
}}}}
%  text 
\put(16845,-10644){\makebox(0,0)[lb]{\smash{{\SetFigFont{25}{30.0}{\familydefault}{\mddefault}{\updefault}{\color[rgb]{0,0,0}$S_2=2/5$}%
}}}}
%  text 
\put(16922,-11107){\makebox(0,0)[lb]{\smash{{\SetFigFont{25}{30.0}{\familydefault}{\mddefault}{\updefault}{\color[rgb]{0,0,0}$b_1=1/10$}%
}}}}
%  text 
\put(16922,-11570){\makebox(0,0)[lb]{\smash{{\SetFigFont{25}{30.0}{\familydefault}{\mddefault}{\updefault}{\color[rgb]{0,0,0}$b_2=3/2$}%
}}}}
%  text 
\put(16884,-12033){\makebox(0,0)[lb]{\smash{{\SetFigFont{25}{30.0}{\familydefault}{\mddefault}{\updefault}{\color[rgb]{0,0,0}$\kappa_1=1/80$}%
}}}}
%  text 
\put(16884,-12497){\makebox(0,0)[lb]{\smash{{\SetFigFont{25}{30.0}{\familydefault}{\mddefault}{\updefault}{\color[rgb]{0,0,0}$\kappa_2=3/10$}%
}}}}
\end{picture}%

%% file: jpdfs_box.pstex_t
\begin{picture}(0,0)%
\includegraphics{jpdfs_box.pstex}%
\end{picture}%
\setlength{\unitlength}{3947sp}%
\begingroup\makeatletter\ifx\SetFigFont\undefined%
\gdef\SetFigFont#1#2#3#4#5{%
  \reset@font\fontsize{#1}{#2pt}%
  \fontfamily{#3}\fontseries{#4}\fontshape{#5}%
  \selectfont}%
\fi\endgroup%
\begin{picture}(19140,18776)(-30,-17937)
%  text 
\put(15120,-10058){\makebox(0,0)[lb]{\smash{{\SetFigFont{25}{30.0}{\familydefault}{\mddefault}{\updefault}{\color[rgb]{0,0,0}$\omega_1=5$}%
}}}}
%  text 
\put(15120,-10483){\makebox(0,0)[lb]{\smash{{\SetFigFont{25}{30.0}{\familydefault}{\mddefault}{\updefault}{\color[rgb]{0,0,0}$\omega_2=2$}%
}}}}
%  text 
\put(15120,-10907){\makebox(0,0)[lb]{\smash{{\SetFigFont{25}{30.0}{\familydefault}{\mddefault}{\updefault}{\color[rgb]{0,0,0}$\omega_3=3$}%
}}}}
%  text 
\put(16998,-10057){\makebox(0,0)[lb]{\smash{{\SetFigFont{25}{30.0}{\familydefault}{\mddefault}{\updefault}{\color[rgb]{0,0,0}$S_1=5/8$}%
}}}}
%  text 
\put(17011,-10521){\makebox(0,0)[lb]{\smash{{\SetFigFont{25}{30.0}{\familydefault}{\mddefault}{\updefault}{\color[rgb]{0,0,0}$S_2=2/5$}%
}}}}
%  text 
\put(17088,-10984){\makebox(0,0)[lb]{\smash{{\SetFigFont{25}{30.0}{\familydefault}{\mddefault}{\updefault}{\color[rgb]{0,0,0}$b_1=1/10$}%
}}}}
%  text 
\put(17088,-11447){\makebox(0,0)[lb]{\smash{{\SetFigFont{25}{30.0}{\familydefault}{\mddefault}{\updefault}{\color[rgb]{0,0,0}$b_2=3/2$}%
}}}}
%  text 
\put(17050,-11910){\makebox(0,0)[lb]{\smash{{\SetFigFont{25}{30.0}{\familydefault}{\mddefault}{\updefault}{\color[rgb]{0,0,0}$\kappa_1=1/80$}%
}}}}
%  text 
\put(17050,-12374){\makebox(0,0)[lb]{\smash{{\SetFigFont{25}{30.0}{\familydefault}{\mddefault}{\updefault}{\color[rgb]{0,0,0}$\kappa_2=3/10$}%
}}}}
%  text 
\put(8067,137){\makebox(0,0)[lb]{\smash{{\SetFigFont{25}{30.0}{\familydefault}{\mddefault}{\updefault}{\color[rgb]{0,0,0}$t=0$}%
}}}}
%  text 
\put(8015,-9135){\makebox(0,0)[lb]{\smash{{\SetFigFont{25}{30.0}{\familydefault}{\mddefault}{\updefault}{\color[rgb]{0,0,0}$t=5$}%
}}}}
%  text 
\put(2958,-8150){\makebox(0,0)[lb]{\smash{{\SetFigFont{25}{30.0}{\familydefault}{\mddefault}{\updefault}{\color[rgb]{0,0,0}0.25}%
}}}}
%  text 
\put(5098,-8150){\makebox(0,0)[lb]{\smash{{\SetFigFont{25}{30.0}{\familydefault}{\mddefault}{\updefault}{\color[rgb]{0,0,0}0.5}%
}}}}
%  text 
\put(7088,-8150){\makebox(0,0)[lb]{\smash{{\SetFigFont{25}{30.0}{\familydefault}{\mddefault}{\updefault}{\color[rgb]{0,0,0}0.75}%
}}}}
%  text 
\put(9336,-8150){\makebox(0,0)[lb]{\smash{{\SetFigFont{25}{30.0}{\familydefault}{\mddefault}{\updefault}{\color[rgb]{0,0,0}1}%
}}}}
%  text 
\put(652,-7776){\makebox(0,0)[lb]{\smash{{\SetFigFont{25}{30.0}{\familydefault}{\mddefault}{\updefault}{\color[rgb]{0,0,0}0}%
}}}}
%  text 
\put(285,-5711){\makebox(0,0)[lb]{\smash{{\SetFigFont{25}{30.0}{\familydefault}{\mddefault}{\updefault}{\color[rgb]{0,0,0}0.25}%
}}}}
%  text 
\put(435,-3646){\makebox(0,0)[lb]{\smash{{\SetFigFont{25}{30.0}{\familydefault}{\mddefault}{\updefault}{\color[rgb]{0,0,0}0.5}%
}}}}
%  text 
\put(285,-1581){\makebox(0,0)[lb]{\smash{{\SetFigFont{25}{30.0}{\familydefault}{\mddefault}{\updefault}{\color[rgb]{0,0,0}0.75}%
}}}}
%  text 
\put(652,484){\makebox(0,0)[lb]{\smash{{\SetFigFont{25}{30.0}{\familydefault}{\mddefault}{\updefault}{\color[rgb]{0,0,0}1}%
}}}}
%  text 
\put(2988,-17390){\makebox(0,0)[lb]{\smash{{\SetFigFont{25}{30.0}{\familydefault}{\mddefault}{\updefault}{\color[rgb]{0,0,0}0.25}%
}}}}
%  text 
\put(5128,-17390){\makebox(0,0)[lb]{\smash{{\SetFigFont{25}{30.0}{\familydefault}{\mddefault}{\updefault}{\color[rgb]{0,0,0}0.5}%
}}}}
%  text 
\put(7118,-17390){\makebox(0,0)[lb]{\smash{{\SetFigFont{25}{30.0}{\familydefault}{\mddefault}{\updefault}{\color[rgb]{0,0,0}0.75}%
}}}}
%  text 
\put(9366,-17390){\makebox(0,0)[lb]{\smash{{\SetFigFont{25}{30.0}{\familydefault}{\mddefault}{\updefault}{\color[rgb]{0,0,0}1}%
}}}}
%  text 
\put(682,-17016){\makebox(0,0)[lb]{\smash{{\SetFigFont{25}{30.0}{\familydefault}{\mddefault}{\updefault}{\color[rgb]{0,0,0}0}%
}}}}
%  text 
\put(315,-14951){\makebox(0,0)[lb]{\smash{{\SetFigFont{25}{30.0}{\familydefault}{\mddefault}{\updefault}{\color[rgb]{0,0,0}0.25}%
}}}}
%  text 
\put(465,-12886){\makebox(0,0)[lb]{\smash{{\SetFigFont{25}{30.0}{\familydefault}{\mddefault}{\updefault}{\color[rgb]{0,0,0}0.5}%
}}}}
%  text 
\put(315,-10821){\makebox(0,0)[lb]{\smash{{\SetFigFont{25}{30.0}{\familydefault}{\mddefault}{\updefault}{\color[rgb]{0,0,0}0.75}%
}}}}
%  text 
\put(682,-8756){\makebox(0,0)[lb]{\smash{{\SetFigFont{25}{30.0}{\familydefault}{\mddefault}{\updefault}{\color[rgb]{0,0,0}1}%
}}}}
%  text 
\put(9852,-1585){\makebox(0,0)[lb]{\smash{{\SetFigFont{25}{30.0}{\familydefault}{\mddefault}{\updefault}{\color[rgb]{0,0,0}0.75}%
}}}}
%  text 
\put(9874,-5672){\makebox(0,0)[lb]{\smash{{\SetFigFont{25}{30.0}{\familydefault}{\mddefault}{\updefault}{\color[rgb]{0,0,0}0.25}%
}}}}
%  text 
\put(17383,-9080){\makebox(0,0)[lb]{\smash{{\SetFigFont{25}{30.0}{\familydefault}{\mddefault}{\updefault}{\color[rgb]{0,0,0}$t=160$}%
}}}}
%  text 
\put(12461,-17382){\makebox(0,0)[lb]{\smash{{\SetFigFont{25}{30.0}{\familydefault}{\mddefault}{\updefault}{\color[rgb]{0,0,0}0.25}%
}}}}
%  text 
\put(14601,-17382){\makebox(0,0)[lb]{\smash{{\SetFigFont{25}{30.0}{\familydefault}{\mddefault}{\updefault}{\color[rgb]{0,0,0}0.5}%
}}}}
%  text 
\put(16591,-17382){\makebox(0,0)[lb]{\smash{{\SetFigFont{25}{30.0}{\familydefault}{\mddefault}{\updefault}{\color[rgb]{0,0,0}0.75}%
}}}}
%  text 
\put(18839,-17382){\makebox(0,0)[lb]{\smash{{\SetFigFont{25}{30.0}{\familydefault}{\mddefault}{\updefault}{\color[rgb]{0,0,0}1}%
}}}}
%  text 
\put(17534, 95){\makebox(0,0)[lb]{\smash{{\SetFigFont{25}{30.0}{\familydefault}{\mddefault}{\updefault}{\color[rgb]{0,0,0}$t=1$}%
}}}}
%  text 
\put(12487,-8095){\makebox(0,0)[lb]{\smash{{\SetFigFont{25}{30.0}{\familydefault}{\mddefault}{\updefault}{\color[rgb]{0,0,0}0.25}%
}}}}
%  text 
\put(14627,-8095){\makebox(0,0)[lb]{\smash{{\SetFigFont{25}{30.0}{\familydefault}{\mddefault}{\updefault}{\color[rgb]{0,0,0}0.5}%
}}}}
%  text 
\put(16617,-8095){\makebox(0,0)[lb]{\smash{{\SetFigFont{25}{30.0}{\familydefault}{\mddefault}{\updefault}{\color[rgb]{0,0,0}0.75}%
}}}}
%  text 
\put(18865,-8095){\makebox(0,0)[lb]{\smash{{\SetFigFont{25}{30.0}{\familydefault}{\mddefault}{\updefault}{\color[rgb]{0,0,0}1}%
}}}}
%  text 
\put(10181,-7721){\makebox(0,0)[lb]{\smash{{\SetFigFont{25}{30.0}{\familydefault}{\mddefault}{\updefault}{\color[rgb]{0,0,0}0}%
}}}}
%  text 
\put(9964,-3591){\makebox(0,0)[lb]{\smash{{\SetFigFont{25}{30.0}{\familydefault}{\mddefault}{\updefault}{\color[rgb]{0,0,0}0.5}%
}}}}
%  text 
\put(10181,539){\makebox(0,0)[lb]{\smash{{\SetFigFont{25}{30.0}{\familydefault}{\mddefault}{\updefault}{\color[rgb]{0,0,0}1}%
}}}}
%  text 
\put(9837,-15011){\makebox(0,0)[lb]{\smash{{\SetFigFont{25}{30.0}{\familydefault}{\mddefault}{\updefault}{\color[rgb]{0,0,0}0.25}%
}}}}
%  text 
\put(9987,-12946){\makebox(0,0)[lb]{\smash{{\SetFigFont{25}{30.0}{\familydefault}{\mddefault}{\updefault}{\color[rgb]{0,0,0}0.5}%
}}}}
%  text 
\put(9837,-10881){\makebox(0,0)[lb]{\smash{{\SetFigFont{25}{30.0}{\familydefault}{\mddefault}{\updefault}{\color[rgb]{0,0,0}0.75}%
}}}}
%  text 
\put(10414,-17068){\makebox(0,0)[lb]{\smash{{\SetFigFont{25}{30.0}{\familydefault}{\mddefault}{\updefault}{\color[rgb]{0,0,0}0}%
}}}}
%  text 
\put(10414,-8808){\makebox(0,0)[lb]{\smash{{\SetFigFont{25}{30.0}{\familydefault}{\mddefault}{\updefault}{\color[rgb]{0,0,0}1}%
}}}}
%  text 
\put(14574,-17810){\makebox(0,0)[lb]{\smash{{\SetFigFont{25}{30.0}{\familydefault}{\mddefault}{\updefault}{\color[rgb]{0,0,0}$Y_1$}%
}}}}
%  text 
\put(9878,-12969){\rotatebox{90.0}{\makebox(0,0)[lb]{\smash{{\SetFigFont{25}{30.0}{\familydefault}{\mddefault}{\updefault}{\color[rgb]{0,0,0}$Y_2$}%
}}}}}
%  text 
\put(9853,-3680){\rotatebox{90.0}{\makebox(0,0)[lb]{\smash{{\SetFigFont{25}{30.0}{\familydefault}{\mddefault}{\updefault}{\color[rgb]{0,0,0}$Y_2$}%
}}}}}
%  text 
\put(284,-3713){\rotatebox{90.0}{\makebox(0,0)[lb]{\smash{{\SetFigFont{25}{30.0}{\familydefault}{\mddefault}{\updefault}{\color[rgb]{0,0,0}$Y_2$}%
}}}}}
%  text 
\put(334,-12953){\rotatebox{90.0}{\makebox(0,0)[lb]{\smash{{\SetFigFont{25}{30.0}{\familydefault}{\mddefault}{\updefault}{\color[rgb]{0,0,0}$Y_2$}%
}}}}}
%  text 
\put(4981,-8485){\makebox(0,0)[lb]{\smash{{\SetFigFont{25}{30.0}{\familydefault}{\mddefault}{\updefault}{\color[rgb]{0,0,0}$Y_1$}%
}}}}
%  text 
\put(5180,-17776){\makebox(0,0)[lb]{\smash{{\SetFigFont{25}{30.0}{\familydefault}{\mddefault}{\updefault}{\color[rgb]{0,0,0}$Y_1$}%
}}}}
%  text 
\put(14567,-8460){\makebox(0,0)[lb]{\smash{{\SetFigFont{25}{30.0}{\familydefault}{\mddefault}{\updefault}{\color[rgb]{0,0,0}$Y_1$}%
}}}}
\end{picture}%

%% file: means.pstex_t
\begin{picture}(0,0)%
\includegraphics{means.pstex}%
\end{picture}%
\setlength{\unitlength}{3947sp}%
\begingroup\makeatletter\ifx\SetFigFont\undefined%
\gdef\SetFigFont#1#2#3#4#5{%
  \reset@font\fontsize{#1}{#2pt}%
  \fontfamily{#3}\fontseries{#4}\fontshape{#5}%
  \selectfont}%
\fi\endgroup%
\begin{picture}(12205,8437)(120,-7651)
%  text 
\put(972,-7263){\makebox(0,0)[lb]{\smash{{\SetFigFont{25}{30.0}{\familydefault}{\mddefault}{\updefault}{\color[rgb]{0,0,0} 0}%
}}}}
%  text 
\put(2431,-7263){\makebox(0,0)[lb]{\smash{{\SetFigFont{25}{30.0}{\familydefault}{\mddefault}{\updefault}{\color[rgb]{0,0,0} 20}%
}}}}
%  text 
\put(3956,-7263){\makebox(0,0)[lb]{\smash{{\SetFigFont{25}{30.0}{\familydefault}{\mddefault}{\updefault}{\color[rgb]{0,0,0} 40}%
}}}}
%  text 
\put(5479,-7263){\makebox(0,0)[lb]{\smash{{\SetFigFont{25}{30.0}{\familydefault}{\mddefault}{\updefault}{\color[rgb]{0,0,0} 60}%
}}}}
%  text 
\put(7004,-7263){\makebox(0,0)[lb]{\smash{{\SetFigFont{25}{30.0}{\familydefault}{\mddefault}{\updefault}{\color[rgb]{0,0,0} 80}%
}}}}
%  text 
\put(8462,-7263){\makebox(0,0)[lb]{\smash{{\SetFigFont{25}{30.0}{\familydefault}{\mddefault}{\updefault}{\color[rgb]{0,0,0} 100}%
}}}}
%  text 
\put(9987,-7263){\makebox(0,0)[lb]{\smash{{\SetFigFont{25}{30.0}{\familydefault}{\mddefault}{\updefault}{\color[rgb]{0,0,0} 120}%
}}}}
%  text 
\put(11511,-7263){\makebox(0,0)[lb]{\smash{{\SetFigFont{25}{30.0}{\familydefault}{\mddefault}{\updefault}{\color[rgb]{0,0,0} 140}%
}}}}
%  text 
\put(660,-6954){\makebox(0,0)[lb]{\smash{{\SetFigFont{25}{30.0}{\familydefault}{\mddefault}{\updefault}{\color[rgb]{0,0,0} 0}%
}}}}
%  text 
\put(465,-5563){\makebox(0,0)[lb]{\smash{{\SetFigFont{25}{30.0}{\familydefault}{\mddefault}{\updefault}{\color[rgb]{0,0,0} 0.1}%
}}}}
%  text 
\put(465,-4173){\makebox(0,0)[lb]{\smash{{\SetFigFont{25}{30.0}{\familydefault}{\mddefault}{\updefault}{\color[rgb]{0,0,0} 0.2}%
}}}}
%  text 
\put(465,-2781){\makebox(0,0)[lb]{\smash{{\SetFigFont{25}{30.0}{\familydefault}{\mddefault}{\updefault}{\color[rgb]{0,0,0} 0.3}%
}}}}
%  text 
\put(465,-1389){\makebox(0,0)[lb]{\smash{{\SetFigFont{25}{30.0}{\familydefault}{\mddefault}{\updefault}{\color[rgb]{0,0,0} 0.4}%
}}}}
%  text 
\put(465,  2){\makebox(0,0)[lb]{\smash{{\SetFigFont{25}{30.0}{\familydefault}{\mddefault}{\updefault}{\color[rgb]{0,0,0} 0.5}%
}}}}
%  text 
\put(10126,-511){\makebox(0,0)[lb]{\smash{{\SetFigFont{29}{34.8}{\familydefault}{\mddefault}{\updefault}{\color[rgb]{0,0,0}$\irmean{Y_1}$}%
}}}}
%  text 
\put(10126,-2461){\makebox(0,0)[lb]{\smash{{\SetFigFont{29}{34.8}{\familydefault}{\mddefault}{\updefault}{\color[rgb]{0,0,0}$\irmean{Y_3}$}%
}}}}
%  text 
\put(10126,-3886){\makebox(0,0)[lb]{\smash{{\SetFigFont{29}{34.8}{\familydefault}{\mddefault}{\updefault}{\color[rgb]{0,0,0}$\irmean{Y_2}$}%
}}}}
%  text 
\put(6151,-7636){\makebox(0,0)[lb]{\smash{{\SetFigFont{29}{34.8}{\familydefault}{\mddefault}{\updefault}{\color[rgb]{0,0,0}time}%
}}}}
%  text 
\put(376,-3736){\rotatebox{90.0}{\makebox(0,0)[lb]{\smash{{\SetFigFont{29}{34.8}{\familydefault}{\mddefault}{\updefault}{\color[rgb]{0,0,0}means}%
}}}}}
\end{picture}%

%% file: secmom.pstex_t
\begin{picture}(0,0)%
\includegraphics{secmom.pstex}%
\end{picture}%
\setlength{\unitlength}{3947sp}%
\begingroup\makeatletter\ifx\SetFigFont\undefined%
\gdef\SetFigFont#1#2#3#4#5{%
  \reset@font\fontsize{#1}{#2pt}%
  \fontfamily{#3}\fontseries{#4}\fontshape{#5}%
  \selectfont}%
\fi\endgroup%
\begin{picture}(24474,8670)(-674,-7801)
%  text 
\put(11788,-7104){\makebox(0,0)[lb]{\smash{{\SetFigFont{25}{30.0}{\familydefault}{\mddefault}{\updefault}{\color[rgb]{0,0,0}-0.12}%
}}}}
%  text 
\put(11917,-5829){\makebox(0,0)[lb]{\smash{{\SetFigFont{25}{30.0}{\familydefault}{\mddefault}{\updefault}{\color[rgb]{0,0,0}-0.1}%
}}}}
%  text 
\put(11788,-4554){\makebox(0,0)[lb]{\smash{{\SetFigFont{25}{30.0}{\familydefault}{\mddefault}{\updefault}{\color[rgb]{0,0,0}-0.08}%
}}}}
%  text 
\put(11788,-3278){\makebox(0,0)[lb]{\smash{{\SetFigFont{25}{30.0}{\familydefault}{\mddefault}{\updefault}{\color[rgb]{0,0,0}-0.06}%
}}}}
%  text 
\put(11788,-2003){\makebox(0,0)[lb]{\smash{{\SetFigFont{25}{30.0}{\familydefault}{\mddefault}{\updefault}{\color[rgb]{0,0,0}-0.04}%
}}}}
%  text 
\put(11788,-728){\makebox(0,0)[lb]{\smash{{\SetFigFont{25}{30.0}{\familydefault}{\mddefault}{\updefault}{\color[rgb]{0,0,0}-0.02}%
}}}}
%  text 
\put(12125,547){\makebox(0,0)[lb]{\smash{{\SetFigFont{25}{30.0}{\familydefault}{\mddefault}{\updefault}{\color[rgb]{0,0,0} 0}%
}}}}
%  text 
\put(165,-7082){\makebox(0,0)[lb]{\smash{{\SetFigFont{25}{30.0}{\familydefault}{\mddefault}{\updefault}{\color[rgb]{0,0,0} 0}%
}}}}
%  text 
\put(-159,-5989){\makebox(0,0)[lb]{\smash{{\SetFigFont{25}{30.0}{\familydefault}{\mddefault}{\updefault}{\color[rgb]{0,0,0} 0.02}%
}}}}
%  text 
\put(-159,-4896){\makebox(0,0)[lb]{\smash{{\SetFigFont{25}{30.0}{\familydefault}{\mddefault}{\updefault}{\color[rgb]{0,0,0} 0.04}%
}}}}
%  text 
\put(-159,-3802){\makebox(0,0)[lb]{\smash{{\SetFigFont{25}{30.0}{\familydefault}{\mddefault}{\updefault}{\color[rgb]{0,0,0} 0.06}%
}}}}
%  text 
\put(-159,-2711){\makebox(0,0)[lb]{\smash{{\SetFigFont{25}{30.0}{\familydefault}{\mddefault}{\updefault}{\color[rgb]{0,0,0} 0.08}%
}}}}
%  text 
\put(-30,-1617){\makebox(0,0)[lb]{\smash{{\SetFigFont{25}{30.0}{\familydefault}{\mddefault}{\updefault}{\color[rgb]{0,0,0} 0.1}%
}}}}
%  text 
\put(-159,-524){\makebox(0,0)[lb]{\smash{{\SetFigFont{25}{30.0}{\familydefault}{\mddefault}{\updefault}{\color[rgb]{0,0,0} 0.12}%
}}}}
%  text 
\put(-159,569){\makebox(0,0)[lb]{\smash{{\SetFigFont{25}{30.0}{\familydefault}{\mddefault}{\updefault}{\color[rgb]{0,0,0} 0.14}%
}}}}
%  text 
\put(12587,-7338){\makebox(0,0)[lb]{\smash{{\SetFigFont{25}{30.0}{\familydefault}{\mddefault}{\updefault}{\color[rgb]{0,0,0} 0}%
}}}}
%  text 
\put(14026,-7338){\makebox(0,0)[lb]{\smash{{\SetFigFont{25}{30.0}{\familydefault}{\mddefault}{\updefault}{\color[rgb]{0,0,0} 20}%
}}}}
%  text 
\put(15531,-7338){\makebox(0,0)[lb]{\smash{{\SetFigFont{25}{30.0}{\familydefault}{\mddefault}{\updefault}{\color[rgb]{0,0,0} 40}%
}}}}
%  text 
\put(17034,-7338){\makebox(0,0)[lb]{\smash{{\SetFigFont{25}{30.0}{\familydefault}{\mddefault}{\updefault}{\color[rgb]{0,0,0} 60}%
}}}}
%  text 
\put(18539,-7338){\makebox(0,0)[lb]{\smash{{\SetFigFont{25}{30.0}{\familydefault}{\mddefault}{\updefault}{\color[rgb]{0,0,0} 80}%
}}}}
%  text 
\put(19977,-7338){\makebox(0,0)[lb]{\smash{{\SetFigFont{25}{30.0}{\familydefault}{\mddefault}{\updefault}{\color[rgb]{0,0,0} 100}%
}}}}
%  text 
\put(21482,-7338){\makebox(0,0)[lb]{\smash{{\SetFigFont{25}{30.0}{\familydefault}{\mddefault}{\updefault}{\color[rgb]{0,0,0} 120}%
}}}}
%  text 
\put(22986,-7338){\makebox(0,0)[lb]{\smash{{\SetFigFont{25}{30.0}{\familydefault}{\mddefault}{\updefault}{\color[rgb]{0,0,0} 140}%
}}}}
%  text 
\put(552,-7316){\makebox(0,0)[lb]{\smash{{\SetFigFont{25}{30.0}{\familydefault}{\mddefault}{\updefault}{\color[rgb]{0,0,0} 0}%
}}}}
%  text 
\put(1991,-7316){\makebox(0,0)[lb]{\smash{{\SetFigFont{25}{30.0}{\familydefault}{\mddefault}{\updefault}{\color[rgb]{0,0,0} 20}%
}}}}
%  text 
\put(3496,-7316){\makebox(0,0)[lb]{\smash{{\SetFigFont{25}{30.0}{\familydefault}{\mddefault}{\updefault}{\color[rgb]{0,0,0} 40}%
}}}}
%  text 
\put(4999,-7316){\makebox(0,0)[lb]{\smash{{\SetFigFont{25}{30.0}{\familydefault}{\mddefault}{\updefault}{\color[rgb]{0,0,0} 60}%
}}}}
%  text 
\put(6504,-7316){\makebox(0,0)[lb]{\smash{{\SetFigFont{25}{30.0}{\familydefault}{\mddefault}{\updefault}{\color[rgb]{0,0,0} 80}%
}}}}
%  text 
\put(7942,-7316){\makebox(0,0)[lb]{\smash{{\SetFigFont{25}{30.0}{\familydefault}{\mddefault}{\updefault}{\color[rgb]{0,0,0} 100}%
}}}}
%  text 
\put(9447,-7316){\makebox(0,0)[lb]{\smash{{\SetFigFont{25}{30.0}{\familydefault}{\mddefault}{\updefault}{\color[rgb]{0,0,0} 120}%
}}}}
%  text 
\put(10951,-7316){\makebox(0,0)[lb]{\smash{{\SetFigFont{25}{30.0}{\familydefault}{\mddefault}{\updefault}{\color[rgb]{0,0,0} 140}%
}}}}
%  text 
\put(2803,-1352){\makebox(0,0)[lb]{\smash{{\SetFigFont{17}{20.4}{\familydefault}{\mddefault}{\updefault}{\color[rgb]{0,0,0}$1/44$}%
}}}}
%  text 
\put(2278,-2864){\makebox(0,0)[lb]{\smash{{\SetFigFont{17}{20.4}{\familydefault}{\mddefault}{\updefault}{\color[rgb]{0,0,0}$21/1100$}%
}}}}
%  text 
\put(2594,-3939){\makebox(0,0)[lb]{\smash{{\SetFigFont{17}{20.4}{\familydefault}{\mddefault}{\updefault}{\color[rgb]{0,0,0}$4/275$}%
}}}}
%  text 
\put(14336,-2759){\makebox(0,0)[lb]{\smash{{\SetFigFont{20}{24.0}{\familydefault}{\mddefault}{\updefault}{\color[rgb]{0,0,0}$-3/550$}%
}}}}
%  text 
\put(14336,-3617){\makebox(0,0)[lb]{\smash{{\SetFigFont{20}{24.0}{\familydefault}{\mddefault}{\updefault}{\color[rgb]{0,0,0}$-1/110$}%
}}}}
%  text 
\put(14336,-5137){\makebox(0,0)[lb]{\smash{{\SetFigFont{20}{24.0}{\familydefault}{\mddefault}{\updefault}{\color[rgb]{0,0,0}$-3/220$}%
}}}}
%  text 
\put(15887,-1586){\makebox(0,0)[lb]{\smash{{\SetFigFont{14}{16.8}{\familydefault}{\mddefault}{\updefault}{\color[rgb]{0,0,0} 0}%
}}}}
%  text 
\put(10501,239){\makebox(0,0)[lb]{\smash{{\SetFigFont{25}{30.0}{\familydefault}{\mddefault}{\updefault}{\color[rgb]{0,0,0}(a)}%
}}}}
%  text 
\put(22426,-736){\makebox(0,0)[lb]{\smash{{\SetFigFont{25}{30.0}{\familydefault}{\mddefault}{\updefault}{\color[rgb]{0,0,0}(b)}%
}}}}
%  text 
\put(17626,-7786){\makebox(0,0)[lb]{\smash{{\SetFigFont{29}{34.8}{\familydefault}{\mddefault}{\updefault}{\color[rgb]{0,0,0}time}%
}}}}
%  text 
\put(5476,-7786){\makebox(0,0)[lb]{\smash{{\SetFigFont{29}{34.8}{\familydefault}{\mddefault}{\updefault}{\color[rgb]{0,0,0}time}%
}}}}
%  text 
\put(2401,-5236){\makebox(0,0)[lb]{\smash{{\SetFigFont{29}{34.8}{\familydefault}{\mddefault}{\updefault}{\color[rgb]{0,0,0}$\irv{y_3}$}%
}}}}
%  text 
\put(1351,-6586){\makebox(0,0)[lb]{\smash{{\SetFigFont{29}{34.8}{\familydefault}{\mddefault}{\updefault}{\color[rgb]{0,0,0}$\irv{y_1}$}%
}}}}
%  text 
\put(6526,-6586){\makebox(0,0)[lb]{\smash{{\SetFigFont{29}{34.8}{\familydefault}{\mddefault}{\updefault}{\color[rgb]{0,0,0}$\irv{y_2}$}%
}}}}
%  text 
\put(13051,-1411){\makebox(0,0)[lb]{\smash{{\SetFigFont{29}{34.8}{\familydefault}{\mddefault}{\updefault}{\color[rgb]{0,0,0}$\irmean{y_2y_3}$}%
}}}}
%  text 
\put(18526,-586){\makebox(0,0)[lb]{\smash{{\SetFigFont{29}{34.8}{\familydefault}{\mddefault}{\updefault}{\color[rgb]{0,0,0}$\irmean{y_1y_3}$}%
}}}}
%  text 
\put(9376,-1411){\makebox(0,0)[lb]{\smash{{\SetFigFont{20}{24.0}{\familydefault}{\mddefault}{\updefault}{\color[rgb]{0,0,0}$\irv{y_1}$}%
}}}}
%  text 
\put(9376,-3286){\makebox(0,0)[lb]{\smash{{\SetFigFont{20}{24.0}{\familydefault}{\mddefault}{\updefault}{\color[rgb]{0,0,0}$\irv{y_2}$}%
}}}}
%  text 
\put(3689,-4903){\makebox(0,0)[lb]{\smash{{\SetFigFont{14}{16.8}{\familydefault}{\mddefault}{\updefault}{\color[rgb]{0,0,0} 80}%
}}}}
%  text 
\put(5798,-4903){\makebox(0,0)[lb]{\smash{{\SetFigFont{14}{16.8}{\familydefault}{\mddefault}{\updefault}{\color[rgb]{0,0,0} 100}%
}}}}
%  text 
\put(7943,-4903){\makebox(0,0)[lb]{\smash{{\SetFigFont{14}{16.8}{\familydefault}{\mddefault}{\updefault}{\color[rgb]{0,0,0} 120}%
}}}}
%  text 
\put(10092,-4903){\makebox(0,0)[lb]{\smash{{\SetFigFont{14}{16.8}{\familydefault}{\mddefault}{\updefault}{\color[rgb]{0,0,0} 140}%
}}}}
%  text 
\put(3376, 14){\makebox(0,0)[lb]{\smash{{\SetFigFont{14}{16.8}{\familydefault}{\mddefault}{\updefault}{\color[rgb]{0,0,0} 0.03}%
}}}}
%  text 
\put(3301,-2311){\makebox(0,0)[lb]{\smash{{\SetFigFont{14}{16.8}{\familydefault}{\mddefault}{\updefault}{\color[rgb]{0,0,0} 0.02}%
}}}}
%  text 
\put(3301,-4636){\makebox(0,0)[lb]{\smash{{\SetFigFont{14}{16.8}{\familydefault}{\mddefault}{\updefault}{\color[rgb]{0,0,0} 0.01}%
}}}}
%  text 
\put(15526,-3961){\makebox(0,0)[lb]{\smash{{\SetFigFont{14}{16.8}{\familydefault}{\mddefault}{\updefault}{\color[rgb]{0,0,0}-0.01}%
}}}}
%  text 
\put(15526,-6286){\makebox(0,0)[lb]{\smash{{\SetFigFont{14}{16.8}{\familydefault}{\mddefault}{\updefault}{\color[rgb]{0,0,0}-0.02}%
}}}}
%  text 
\put(15976,-6436){\makebox(0,0)[lb]{\smash{{\SetFigFont{14}{16.8}{\familydefault}{\mddefault}{\updefault}{\color[rgb]{0,0,0} 80}%
}}}}
%  text 
\put(18001,-6436){\makebox(0,0)[lb]{\smash{{\SetFigFont{14}{16.8}{\familydefault}{\mddefault}{\updefault}{\color[rgb]{0,0,0} 100}%
}}}}
%  text 
\put(20101,-6436){\makebox(0,0)[lb]{\smash{{\SetFigFont{14}{16.8}{\familydefault}{\mddefault}{\updefault}{\color[rgb]{0,0,0} 120}%
}}}}
%  text 
\put(22201,-6436){\makebox(0,0)[lb]{\smash{{\SetFigFont{14}{16.8}{\familydefault}{\mddefault}{\updefault}{\color[rgb]{0,0,0} 140}%
}}}}
%  text 
\put(21376,-2611){\makebox(0,0)[lb]{\smash{{\SetFigFont{20}{24.0}{\familydefault}{\mddefault}{\updefault}{\color[rgb]{0,0,0}$\irmean{y_2y_3}$}%
}}}}
%  text 
\put(21376,-3511){\makebox(0,0)[lb]{\smash{{\SetFigFont{20}{24.0}{\familydefault}{\mddefault}{\updefault}{\color[rgb]{0,0,0}$\irmean{y_1y_2}$}%
}}}}
%  text 
\put(21376,-4561){\makebox(0,0)[lb]{\smash{{\SetFigFont{20}{24.0}{\familydefault}{\mddefault}{\updefault}{\color[rgb]{0,0,0}$\irmean{y_1y_3}$}%
}}}}
%  text 
\put(9376,-2236){\makebox(0,0)[lb]{\smash{{\SetFigFont{20}{24.0}{\familydefault}{\mddefault}{\updefault}{\color[rgb]{0,0,0}$\irv{y_3}$}%
}}}}
%  text 
\put(14851,-961){\makebox(0,0)[lb]{\smash{{\SetFigFont{29}{34.8}{\familydefault}{\mddefault}{\updefault}{\color[rgb]{0,0,0}$\irmean{y_1y_2}$}%
}}}}
%  text 
\put(11626,-4186){\rotatebox{90.0}{\makebox(0,0)[lb]{\smash{{\SetFigFont{29}{34.8}{\familydefault}{\mddefault}{\updefault}{\color[rgb]{0,0,0}covariances}%
}}}}}
%  text 
\put(-299,-4111){\rotatebox{90.0}{\makebox(0,0)[lb]{\smash{{\SetFigFont{29}{34.8}{\familydefault}{\mddefault}{\updefault}{\color[rgb]{0,0,0}variances}%
}}}}}
\end{picture}%